\documentclass[twocolumn,pre,showpacs,amsmath]{revtex4-1}

\usepackage[dvipdfmx]{graphicx}

\newcommand{\rd}[3][]{\frac{\partial^{#1} #2}{{\partial #3}^{#1}}}
\newcommand{\rdt}[3][]{{\partial^{#1} #2}/{{\partial #3}^{#1}}}
\newcommand{\dt}[1]{\frac{\mathrm d #1}{\mathrm dt}}
\newcommand{\dtt}[1]{{\mathrm d #1}/{\mathrm dt}}

\newcommand{\lt}[1][t]{\lambda_{#1}}

\newcommand{\tsratio}{\eta}

\newcommand{\noneq}{\epsilon}

\newcommand{\Sh}{S}
\newcommand{\Hs}{\Sh_{\rm sym}}
\newcommand{\Pt}[1][t]{P_{#1}}
\newcommand{\Pm}[1][x]{\Pt({#1})}
\newcommand{\td}{{t + \Delta t}}
\newcommand{\Llt}{\mathcal{L}_{\lt}}
\newcommand{\Pss}[2]{P_{\rm ss}^{#1}({#2})}
\newcommand{\conj}[1]{\bar{#1}}

\newcommand{\sbath}[2][\lt]{\sigma[#2; #1]}
\newcommand{\ssbath}[2]{\hat\sigma(#1 \rightarrow #2, \lt)}
\newcommand{\sirr}[1]{\Sigma[#1; \lt]}

\newcommand{\sKNSTex}[2][\lt]{\sigma_{\rm ex}[#2; #1]}

\newcommand{\X}{\Gamma}

\newcommand{\res}[2][]{R^{#1}[#2]}

\begin{document}

\title{Invariance of Steady State Thermodynamics between Different Scales of Description}
\author{Yohei Nakayama}
 \email[]{nakayama@daisy.phys.s.u-tokyo.ac.jp}
\author{Kyogo Kawaguchi}
 \email[]{kyogok@daisy.phys.s.u-tokyo.ac.jp}
\affiliation{Department of Physics, The University of Tokyo, Bunkyo, Tokyo 113-0033, Japan}
\date{\today}

\begin{abstract}
By considering general Markov stochastic dynamics and its coarse-graining,
we study the framework of stochastic thermodynamics for the original and reduced descriptions
corresponding to different scales.
We are especially concerned with the case where the irreversible entropy production has a finite difference between the scales.
We find that the sum of increment of nonequilibrium entropy and excess part of entropy production, which are key quantities in construction of steady state thermodynamics, is essentially kept invariant with respect to the change in the scales of description.
This general result justifies experimental approaches toward steady state thermodynamics based on coarse-grained variables.
We demonstrate our result in a mesoscopic heat engine system.
\end{abstract}

\pacs{
05.40-a 
, 05.70.Ln 
}

\maketitle

\section{Introduction}
Recent developments in experimental techniques enable the manipulation of small systems and the measurement of fluctuations therein.
In the framework of stochastic thermodynamics,
 thermodynamic quantities such as heat, work and entropy, are expressed in terms of kinetic equations that the stochastic dynamics of the system follows.
The mesoscopic version of thermodynamic relations between these quantities may be verified and utilized in various experimental setups including biomolecules \cite{liphardt_equilibrium_2002}, colloidal particle systems \cite{wang_experimental_2002},
molecular motors \cite{toyabe_nonequilibrium_2010}, and mesoscopic electric systems \cite{saira_test_2012}.

When one applies the framework of stochastic thermodynamics to the analysis of experimental results,
we should be aware that
models at different time or length scales,
may describe the dynamics of the same physical system.
Although different descriptions are meant to present the same dynamics in an appropriate limit,
it has been shown that even in such case,
 naive application of stochastic thermodynamics may lead to different results between different scales of description \cite{sekimoto_stochastic_2010}.

The scale-dependence of irreversible entropy production has recently attracted particular attention due to
its influence on the second law of thermodynamics.
Since the irreversible entropy production is the sum of the entropy production in the heat bath(s) and the entropy increment in the system, the dependence of its magnitude on the scales of description indicates that the relation between thermodynamic quantities is not invariant.
Furthermore, the dependence affects the discussion of the efficiency of heat engines.
For example,
a mesoscopic heat engine driven by imposing spatial modulations of temperature and potential on a Brownian particle system \cite{buttiker_transport_1987},
 has a finite difference in the irreversible entropy production rate between the underdamped and overdamped descriptions,
causing the second law to have totally different forms depending on the scales.
This means that the efficiency of such heat engine could only be described in a scale-dependent manner \cite{hondou_unattainability_2000}, contrary to our expectation that the fundamental thermodynamic nature would be described in an objective way.
This problem has also been investigated in the context of effect of coarse-graining on the entropy balance \cite{esposito_stochastic_2012}, anomaly due to a symmetry-breaking \cite{celani_anomalous_2012} and the origin of irreversibility \cite{kawaguchi_fluctuation_2013}.

In this paper, 
we take a different approach to overcome the problem of this scale-dependence of the irreversible entropy production.
The key concept is to find some scale-independent quantity which 
plays a role similar to the entropy production.
We focus on the excess part of the entropy production,
defined by subtracting the steady state housekeeping part from the entropy production
 in the context of steady state thermodynamics, a possible extension of thermodynamics to nonequilibrium steady states \cite{landauer_dqt_1978,oono_steady_1998}.
Although the excess entropy production we are mainly concerned with has been introduced in order to
determine a thermodynamic potential through quasi-static operations \cite{komatsu_steady-state_2008,komatsu_entropy_2010},
we find, for general physical systems modeled by Markov processes, that the sum of the excess entropy production \cite{komatsu_steady-state_2008} and the system's entropy increment
may be kept essentially invariant with respect to the change in the scales of description,
even in the case of non-quasistatic operation.
Since the invariance of excess entropy productions implies that nonequilibrium versions of the second law may have scale-independent descriptions,
this may pave the way toward verification of these relations in the realistic experimental situation, where the measurable quantities are often restricted to the coarse-grained scale.

\section{Setup} 
Let us consider a system where slow degrees of freedom, \(x\), and fast degrees of freedom, \(y\), interact with each other and with environment.
Due to the effect of the environment, the time evolution of \((x, y)\) may follow a stochastic dynamics.
We assume this stochastic dynamics to be Markovian, which is justified if the time scale of the equilibration of the environment is much shorter than that of the system.

The Markovian stochastic process of \((x,y)\) is described by the master equation that the probability density function \(\Pm[x,y]\) follows,
\begin{align}
 \rd{\Pt(x,y)}{t} = \Llt \cdot \Pm[x,y]
 .
 \label{e: original master equation}
\end{align}
Here, \(\Llt\) is the generator of time evolution, whose subscript represents implicit dependence on time through controllable parameter(s), \(\lt\).
\(\lt\) may include, for example, the volume of the system, the position or strength of an optical tweezer, temperature(s) of the environment, and the strength of non-conservative force.
Thermodynamic operations are performed through \(\lt\).

Next, we consider the stochastic process of \(x\) obtained from that of \((x,y)\) by tracing out \(y\).
This stochastic process corresponds to the view at the coarse-grained scale, where we may observe only \(x\).
This reduced (coarse-grained) dynamics is not always Markovian, since the probability distribution of \(y\), in general, also depends on the values of \(x\) at the earlier times.
However, if the dynamics of \(y\) is sufficiently faster than that of \(x\) and the modulation of \(\lt\),
then \(\Pt(x,y)\) following Eq.~(\ref{e: original master equation}) typically converges to
\begin{equation}
 \Pm[x,y] = P^{\lt}(y|x) \Pm[x] + O(\tsratio)
 ,
 \label{e:slaving}
\end{equation}
in the longer time scale than \(\tau_y\), and the Markov property of \(x\) is justified.
Here, we define the time scales of \(x,y\) and \(\lt\) as \(\tau_x, \tau_y\) and \(\tau_\lambda\), respectively, and the separation of time scales, \(\tsratio := \max\{\tau_y / \tau_x, \tau_y / \tau_\lambda\}\), which serves as a small parameter.
Equation (\ref{e:slaving}) may be interpreted as the fast relaxation of \(y\) to the local steady density function, \(P^{\lt}(y|x)\), under the given \(x\) and \(\lt\).

\section{Stochastic thermodynamics}
We give a brief review of stochastic thermodynamics in the following paragraph \cite{seifert_stochastic_2012}.
For several systems (including \cite{liphardt_equilibrium_2002, wang_experimental_2002, toyabe_nonequilibrium_2010, saira_test_2012}), the entropy production in the framework of stochastic thermodynamics may be written in a unified form. 
\(\X\) generically denotes the variable(s) of the stochastic dynamics, such as \(x\) and \(y\).

First, we consider the entropy production rate in the heat baths, \(\sigma\), which is the sum of the energy currents flowing from the system to the baths divided by the respective temperatures of the baths. The entropy production rate when the variable(s) of system changes from \(\X\) to \(\X'\), is related to
 the transition probability from \(\X\) to \(\X'\) between time \(t\) and \(\td\), \(W^{\lt}_{\Delta t}(\X'|\X)\), as
\begin{align}
 \ssbath{\X}{\X'} := \lim_{\Delta t\rightarrow0}\frac{1}{\Delta t} \ln \frac{W^{\lt}_{\Delta t}(\X'|\X)}{W^{\lt}_{\Delta t}(\conj\X|\conj\X')}
 .
 \label{e:sSinbaths}
\end{align}
\(\conj\X\) indicates the time-reversal of \(\X\),
and the Boltzmann constant is set to unity throughout this paper.
The entropy production rate upon a given state \(\Pm[\X]\)
is obtained as the expected value of the entropy production rate [Eq.~(\ref{e:sSinbaths})],
\begin{align}
 \sbath{\Pm[\X]} := \lim_{\Delta t\rightarrow0}\langle \ssbath{\X_t}{\X_\td}\rangle_{\Pt[\td, t](\X_\td, \X_t)}
 ,
 \label{e:Sinbaths}
\end{align}
with respect to the joint probability density function \(\Pt[\td, t](\X_\td, \X_t) = W^{\lt}_{\Delta t}(\X_\td|\X_t)\Pm[\X_t]\).
Next, the entropy of the system at time \(t\) is given by the Shannon entropy of \(\Pm[\X]\),
\begin{align}
 \Sh[\Pm[\X]] := - \int \Pm[\X] \ln \Pm[\X] \mathrm d\X.
 \label{e:Shannon}
\end{align}
Then, we may prove the second law of thermodynamics,
\begin{align}
 \Sh[\Pt[\mathcal T](\X)] - \Sh[\Pt[0](\X)] \geq - \int^{\mathcal T}_0 \sbath{\Pm[\X]} \mathrm dt
 ,
 \label{e:second law}
\end{align}
for an arbitrary initial probability density function \(\Pt[0](\X)\) and protocol \((\lt)_{t=0}^\mathcal{T}\).
The equality of Eq.~(\ref{e:second law}) is achieved when
the system is initially in the equilibrium state for \(\lt[0]\),
and the controllable parameter \(\lt\) is slowly varied under the condition that each \(\lt\) corresponds to an equilibrium state.
We call the sum,
\begin{align}
\sirr{\Pm[\X]} := \dt{\Sh[\Pm[\X]]} + \sbath{\Pm[\X]},
\end{align}
the irreversible entropy production rate, which should be interpreted as the entropy production rate in the total system.

Contrary to the expectation that these expressions [Eqs.~(\ref{e:Sinbaths},\ref{e:Shannon})] are consistent with the thermodynamically-defined entropy productions in various systems,
they do not always work well, especially when a non-equilibrium system is subjected to coarse-graining.
We may show in some examples \cite{esposito_stochastic_2012, celani_anomalous_2012, kawaguchi_fluctuation_2013} that the quantities \(\sbath{\Pm[\X]}, \Sh[\Pm[\X]]\) and \(\sirr{\Pm[\X]}\) defined using the fine- and coarse-grained variables [by Eqs.~(\ref{e:Sinbaths},\ref{e:Shannon})] have finite differences,
even in the limit of \(\tsratio\rightarrow0\),
where the Markov property of the coarse-grained description is completely satisfied.
The differences in these quantities mean that Eqs.~(\ref{e:Sinbaths},\ref{e:Shannon}) fail to give thermodynamically-defined entropy productions, either at the fine- or coarse-grained description.
In general, the expressions of the thermodynamic entropy productions cannot be derived solely from the mathematical description of the process, but one needs to take into account how the description is obtained from physical system.

\section{Invariance of excess entropy production} 
Our main aim is to investigate the dependence of entropy-like quantities on the scales of description.
The key quantity we consider is the excess entropy production.
The excess entropy production has been introduced to extend the thermodynamic structure to the nonequilibrium states \cite{landauer_dqt_1978,oono_steady_1998}.
In the nonequilibrium steady states, unlike the equilibrium states, the entropy of the total system is produced steadily and diverges in time.
As a consequence, the straightforward application of the second law of thermodynamics does not give any meaningful inequality.
Since the excess entropy production is defined by subtracting the steady state housekeeping entropy production from the entropy production,
it is possible to derive some useful relations by replacing the entropy production in the second law with its excess part.

In \cite{komatsu_steady-state_2008}, Komatsu {\it et al.} identified the housekeeping entropy production rate at time \(t\) with the entropy production rate upon steady state, \(\Pss{\lt}{\X}\), with the parameter fixed at \(\lt\):
\(\sbath[\lt]{\Pss{\lt}{\X}}\).
The excess entropy production rate in the heat baths is given as
\begin{equation}
 \sKNSTex[\lt]{\Pm[\X]} = \sbath[\lt]{\Pm[\X]} - 
 \sbath[\lt]{\Pss{\lt}{\X}}.
 \label{e:excess}
\end{equation}
In the quasistatic change of
parameter(s) from \(\lt[0]\) to \(\lt[\mathcal T]\) causing
a transition between the steady states \(\Pss{\lt[0]}{\X}\) and \(\Pss{\lt[\mathcal{T}]}{\X}\),
\(\sKNSTex[\lt]{\Pm[\X]}\) satisfies the extended Clausius relation \cite{komatsu_steady-state_2008},
\begin{align}
 \Hs[\Pt[\mathcal T](\X)] - \Hs[\Pt[0](\X)]=& -\int_0^{\mathcal T} \! \sKNSTex{\Pt(\X)} \mathrm dt
 + \res{\lt}
 .
 \label{eq: extended Clausius}
\end{align}
Equation (\ref{eq: extended Clausius}) states that the change in the symmetrized Shannon entropy,
\begin{align}
 \Hs[\Pm[\X]] := -\int \mathrm d\X \Pm[\X] \frac{1}{2}\left[\ln \Pm[\X] + \ln \Pm[\conj\X]\right],
\end{align}
is equal to the excess entropy production within a residual error term, \(\res{\lt}\).
Since \(\sirr{\Pm[\X]} \geq 0\), the lowest order of \(\res{\lt}\) is \(O(\noneq^2)\), where \(\noneq\) denotes the degree of nonequilibrium such as the temperature difference between two heat baths divided by the mean temperature.
However, Komatsu {\it et al.} showed that \(\res{\lt}\)
 can be decreased to \(O(\noneq^3)\) by taking an appropriate protocol linking the given initial and final steady states \cite{komatsu_steady-state_2008}.
Hereafter, we restrict ourselves to such appropriate protocols which achieve \(\res{\lt} = O(\noneq^3)\) under quasi-static operations, and examine the case of finite-time operations, which is beyond the original theory \cite{komatsu_steady-state_2008}.

We find even for non-quasistatic parameter change between the steady states that
the sum of the symmetrized Shannon entropy increment and the excess entropy production,
\begin{align}
 &\Hs[\Pt[\mathcal T](\X)] - \Hs[\Pt[0](\X)] + \int_0^\mathcal{T} \sKNSTex{\Pm[\X]} \mathrm dt
 \nonumber \\
 &=: \Omega[\Pt[0]{(\X)}, \lt],
 \label{e:definition of invariant}
\end{align}
is kept invariant within errors of \(\res[\rm qs]{\Pt[0]{(x)}, \lt}\), \(\res[\rm nqs]{\Pt[0]{(x)}, \lt}\) and \(O(\tsratio)\) as,
\begin{align}
 &\Omega[\Pt[0](x,y), \lt] 
 = \Omega[\Pt[0](x), \lt] 
 \nonumber
 \\
 &+ \res[\rm qs]{\Pt[0]{(x)}, \lt} + \res[\rm nqs]{\Pt[0]{(x)}, \lt} + O(\tsratio) + O(\noneq^3),
 \label{e:invariance of KNST}
\end{align}
for the case where Eq.~(\ref{e:slaving}) holds and 
\begin{flalign}
 &	&	P^{\lt}(y|x) &= P^{\lt}(\conj y|\conj x) + O(\noneq^2)
 \label{e:symmetry condition}
 \\
 &\mbox{or}&	\Pm[x] &= \Pm[\conj{x}] + O(\noneq) &&
 \label{e:overdamped condition}
\end{flalign}
is satisfied.
We introduce the excess entropy production rates to the \((x,y)\) and \(x\) scales of description, \(\sKNSTex{\Pm[x,y]}\) and \(\sKNSTex{\Pm[x]}\), from the entropy production rates \(\sbath{\Pm[x,y]}\) and \(\sbath{\Pm[x]}\), respectively, following their definition [Eq.~(\ref{e:excess})].
\(\res[\rm qs]{\Pt[0]{(x)}, \lt}\) is of the same order as \(R\) for the corresponding quasistatic operation, 
meaning that now \(\res[\rm qs]{\Pt[0]{(x)}, \lt} = O(\noneq^3)\).
The major difference in the residual error from the extended Clausius relation [Eq.~(\ref{eq: extended Clausius})] is its dependence on the speed of operation, which is represented by the \(\res[\rm nqs]{\Pt[0]{(x)}, \lt}\) term.
Here, \(\res[\rm nqs]{\Pt[0]{(x)}, \lt}\) is an accumulated change of the deviation of the coarse-grained probability density function from the steady state one with the parameter fixed at \(\lt\),
which could be evaluated at the coarse-grained scale.
As we shall see in the following, it can also be decreased to \(O(\noneq^3)\) even when the operation is fast. This means that the difference in \(\Omega\) between the scales is not necessarily large in the sense of the extended Clausius relation.
Derivation and the definitions of \(\res[\rm qs]{\Pt[0]{(x)},\lt}\) and \(\res[\rm nqs]{\Pt[0]{(x)}, \lt}\) are given in Appendix \ref{app}.

Among the conditions for Eq.~(\ref{e:invariance of KNST}) to hold, Eq.~(\ref{e:symmetry condition}) and (\ref{e:overdamped condition}) mean the time-reversal symmetry of the local steady density function \(P^{\lt}(y|x)\) and that of the coarse-grained density function \(\Pm[x]\), respectively.
Therefore, Eq.~(\ref{e:symmetry condition}) [or Eq.~(\ref{e:overdamped condition})] trivially holds when \((x,y)\) (or \(x\)) does not include the time-reversal antisymmetric variables such as momentum.
We note that Eqs.~(\ref{e:symmetry condition}) and (\ref{e:overdamped condition}) may be replaced by a conditon \(\rdt{P^\lambda(y|x)}{\lambda} = O(\noneq)\).
However, this condition is formal and hardly ever satisfied in a realistic situation.
Note that when Eq.~(\ref{e:symmetry condition}) holds without its error term, fluctuation theorem is derived for the difference in the irreversible entropy production, \(\Sigma\) \cite{kawaguchi_fluctuation_2013}.

Although there exist different types of excess entropy production corresponding to different definitions of the housekeeping part,
we find the nonequilibrium thermodynamic relations to be universally kept invariant with respect to the change in the scales of description.
In the case of the excess entropy production originally defined by Hatano and Sasa \cite{hatano_steady-state_2001},
we may extend the statement by Santill\'an and Qian \cite{santillan_irreversible_2011} and show for any Markovian stochastic dynamics satisfying Eq.~(\ref{e:slaving}) that the sum of the Shannon entropy increment and the Hatano-Sasa excess entropy production is kept invariant.
It may be shown for another definition introduced by Maes and Neto\v cn\'y \cite{maes_nonequilibrium_2014} that the same invariance holds in a two dimensional overdamped Langevin model example.
The details will be discussed elsewhere.

\section{Examples}
In the following, we consider 
the B\"uttiker-Landauer motor system, where
spatial modulations of temperature and potential are imposed on a Brownian particle
\cite{buttiker_transport_1987}.
The motion of Brownian particle under such a situation may be described by an underdamped Langevin equation with the position-dependent temperature, \(T(x)\),
\begin{align}
  \dot x &= \frac{p}{m}
  , &
  \dot p &= -\gamma \frac{p}{m} - \rd{U(x)}{x}+ \sqrt{2\gamma T(x)}\xi
  .
 \label{e:BL underdamped}
\end{align}
Here, \(x\) and \(p\) are the position and momentum of the particle, respectively,
\(m\) is the mass of the particle, \(\gamma\) is the drag coefficient, \(U(x)\) is the mechanical potential, and \(\xi\) is a white Gaussian noise with zero mean and unit variance.
A periodic boundary condition with period \(L\) is imposed on \(x\).
We treat \(U(x), T(x)\) as the controllable parameters depending on \(t\).

In this case, \((x,p)\) corresponds to \((x,y)\) of our general setup and \(\tsratio = \max\{\tau_p/\tau_x, \tau_p/\tau_\lambda\}\) characterizes the separation of time scales, where \(\tau_p := m/\gamma\).
Applying the standard procedure of the singular perturbation theory \cite{van_kampen_stochastic_1992}, we obtain the asymptotic behavior of the probability density function at time scale longer than \(\tau_p\),
\begin{align}
 \Pm[x,y] = \Pm[x] \frac{1}{\sqrt{2\pi mT(x)}} e^{-\frac{p^2}{2mT(x)}}+ O(\tsratio)
 ,
 \label{e:BL asymptotic}
\end{align}
and an overdamped Langevin equation that \(x\) follows at the coarse-grained scale,
\begin{align}
 \gamma \dot x = -\rd{U(x)}{x} - \frac{1}{2}\rd{T(x)}{x} + \sqrt{2\gamma T(x)}\circ\xi
 .
 \label{e:BL overdamped}
\end{align}
Here and in what follows, we omit the higher order terms [\(O(\tsratio)\)].
The symbol \(\circ\) denotes the product in the Stratonovich sense.
By substituting Eq.~(\ref{e:BL asymptotic}) into Eqs.~(\ref{e:Sinbaths}) and (\ref{e:Shannon}),
we may obtain the asymptotic expression of \(\sirr{\Pm[x,p]}\) as
\begin{align}
 \sirr{\Pm[x,p]} = \sirr{\Pm[x]} + \left\langle \frac{T(x)}{2\gamma} \left(\frac{1}{T(x)}\rd{T(x)}{x}\right)^2\right\rangle
 ,
 \label{e:hidden}
\end{align}
which indicates the presence of finite difference in the irreversible entropy production, \(\Sigma\), except for the isothermal cases \cite{celani_anomalous_2012}.
In contrast, the invariance of \(\Omega\) holds since Eq.~(\ref{e:BL asymptotic}) satisfies Eqs.~(\ref{e:slaving}) and (\ref{e:symmetry condition}).

\begin{figure}[tbp]
 \includegraphics[width=\hsize, clip]{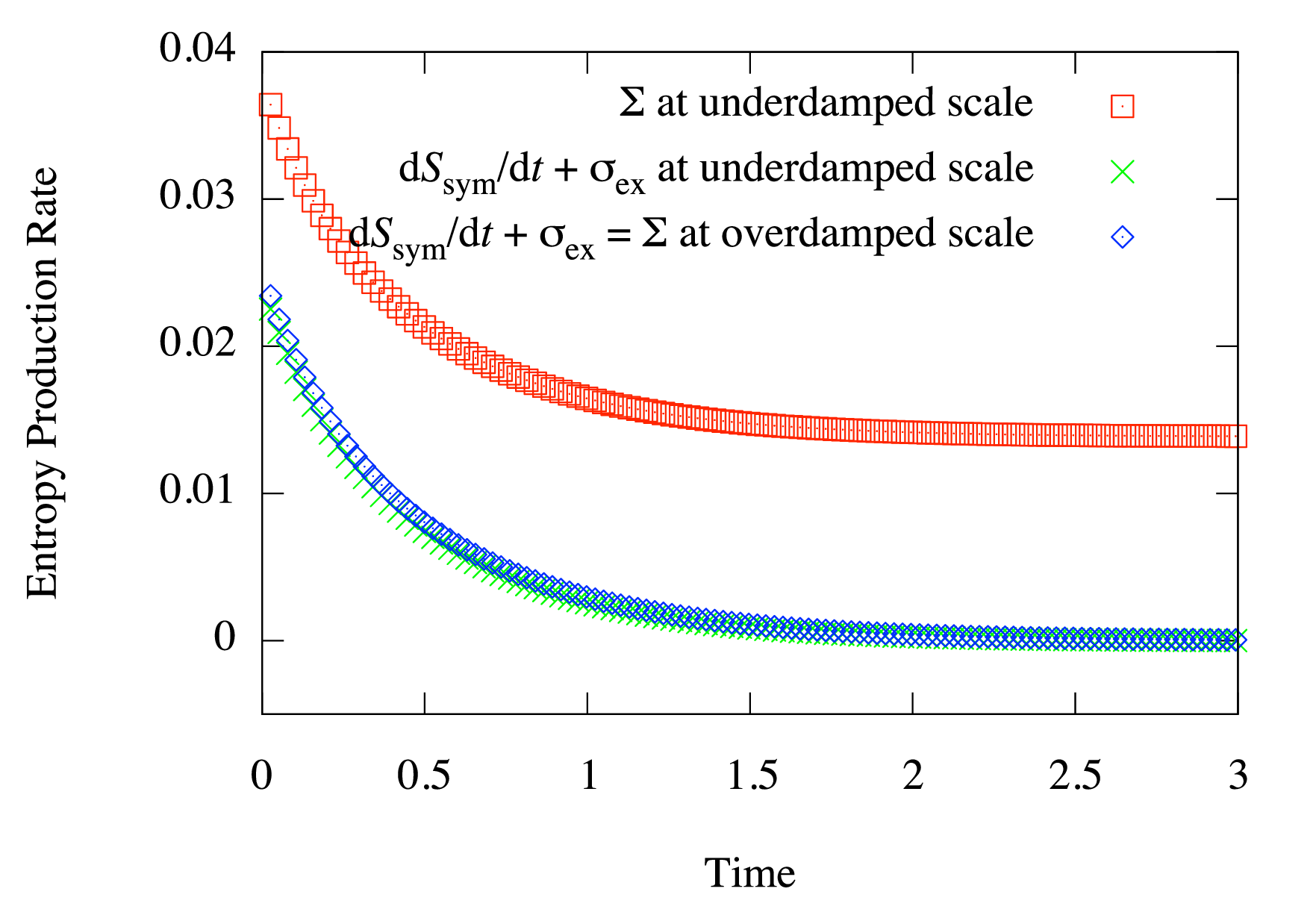}
 \caption{(Color online) The time series of the entropy production rates for the example [Eq.~(\ref{e:BL underdamped})].
 The plotted quantities are \(\sirr{\Pm[\X]}\) at the underdamped scale (square, red), at the overdamped scale (diamond, blue) and \(\dtt{\Hs[\Pm[\X]]} + \sKNSTex{\Pm[\X]}\) at the underdamped scale (cross, green). \(\noneq\) is set as \(0.1\). The time scale, \({L^2 \gamma}/{4 \pi^2 T_0}\), is taken as the unit of time.
 }
 \label{f:numerical}
\end{figure}

Let us now consider a sudden quench of the controllable parameters from \(\lt[{\rm i}] := \left(U_{\rm i}(x) = \noneq\phi(x), T_{\rm i} = T_0\right)\) to \(\lt[{\rm f}] := \left(U(x) = U_{\rm i}(x), T(x) = T_0 - 2\noneq\phi(x)\right)\) at time \(t = 0\).
By taking \(\phi(x) = T_0 [\cos(2\pi x/L) + 1/4 \cos(4\pi x/L)]\),
the probability density function after the quench is given up to \(O(\noneq)\) as
\begin{align}
 \Pm =& \Pss{\lt[{\rm f}]}{x} - 2\noneq \biggl[e^{-\left(\frac{2\pi}{L'}\right)^2 \frac{T_0}{\gamma} t} \cos\left(\frac{2\pi x}{L}\right)
 \nonumber
 \\
 &+ \frac{1}{4} e^{-\left(\frac{4\pi}{L'}\right)^2 \frac{T_0}{\gamma} t} \cos\left(\frac{4\pi x}{L}\right) \biggr] + O(\noneq^2)
 \label{e:Pt}
 ,
\end{align}
where \(L' := \int_0^L \sqrt{{T_0}/{T(x)}} \mathrm dx\).
Equation~(\ref{e:Pt}) agrees with the initial equilibrium and final steady density function,
\begin{align}
 		\Pss{\lt[{\rm i}]}{x} &\propto \exp[-\noneq\phi(x)/T_0]
 &\mbox{and}&&	\Pss{\lt[{\rm f}]}{x} &= \sqrt{\frac{T_0}{T(x)}} \frac{1}{L'}
 , 
\end{align}
at \(t = 0\) and \(t\rightarrow\infty\), respectively.
\(\noneq\) represents the degree of nonequilibrium after the quench, hence \(\noneq \propto \max\{\left|L T'(x)/(2\pi T_0)\right|\} \).
In this example, it immediately follows from \(\Pm - \Pss{\lt[{\rm f}]}{x} = O(\noneq)\) that \(\res[\rm nqs]{\Pt[0]{(x)}, \lt} = O(\noneq^3)\). Therefore, we find the difference in \(\Omega\) [Eq.~(\ref{e:definition of invariant})] between the scales to be \(O(\noneq^3)\),
which may be also calculated explicitly as
\begin{align}
 \Omega[P_0(x,p), \lt] - \Omega[P_0(x), \lt] = -\frac{15}{16}
 \noneq^3 + O(\noneq^4)
 ,
 \label{e:two-mode result}
\end{align}
where we take the ensemble average of Eq.~(\ref{e:hidden}) with respect to \(P_t(x)\) and \(\Pss{\lt[{\rm f}]}{x}\), and integrate it over \(t\in[0,\mathcal{T}]\).
In contrast to the difference in \(\Omega\) [Eq.~(\ref{e:two-mode result})], the difference in the irreversible entropy production \(\int_0^\mathcal{T} \Sigma \mathrm dt\) between the scales [refer to Eq.~(\ref{e:hidden})] includes an obviously large contribution diverging with the elapsed time after quench, \(\mathcal{T}\), as \(\mathcal{T} \int_0^L {\mathrm dx}/{L'} {\sqrt{T_0 T(x)}}/{\gamma}\left[{T'(x)}/{T(x)}\right]^2\).
In this model, we may easily calculate \(\Omega[P_0(x), \lt]\) by using the correspondence of the overdamped dynamics to an appropriate equilibrium dynamics, as
\begin{align}
 \Omega[P_0(x), \lt]= \int \mathrm dx \Pss{\lt[{\rm i}]}{x} \ln \frac{\Pss{\lt[{\rm i}]}{x}}{\Pss{\lt[{\rm f}]}{x}} = O(\noneq^2) 
 .
 \label{e:KL}
\end{align}
This means that the extended Clausius relation (\ref{eq: extended Clausius}) is not satisfied up to \(O(\noneq^3)\) at both scales of description.
Therefore, this model serves as an example of the case where \(\Omega\) is large, but is invariant with respect to the change in the scales of description.
In \figurename~\ref{f:numerical}, we show the numerical result confirming that the difference in \(\Omega\) between the scales may be small
even when the difference in the irreversible entropy production, \(\Sigma\), between the scales and \(\Omega\) itself are large.

\section{Conclusion}
We have considered the system which can be described as Markovian stochastic process at two scales of description, and have shown that the excess entropy production is essentially invariant with respect to the change in the scales of description, even in the case where the irreversible entropy production depends on the scales and the operation is not quasistatic.
The main result has been illustrated in B\"uttiker-Landauer motor system.

Our results are encouraging for the experimental investigation of steady state thermodynamics.
Measurable variables are often restricted to a few degrees of freedom,
due to the limitation in experimental techniques.
In such cases, the irreversible entropy production does not generally have a scale-invariant value.
It is expected, however, that the excess entropy production possesses a scale-independent value,
when we obtain the coarse-grained description of the measurable quantities.
Therefore, the objectivity of the physical quantity is recovered by considering the excess entropy production, supporting the possibility of experimentally exploring steady state thermodynamics based on the mesoscopic scale of description.

We are grateful to S-i. Sasa, Y. Izumida, M. Sano, K. A. Takeuchi and T. Sagawa for fruitful discussions.
This work is supported by the Grant-in-Aids for JSPS Fellow (Grants No. 24-3258 and No. 24-8031).

\appendix

\newcommand{\cSbath}[2]{\langle\Theta_{(#2)}\rangle^{(#2)}_{#1\rightarrow{\rm ss}}}
\newcommand{\cTheta}[2][*]{\cSbath{\Gamma^{#1}}{#2}}
\renewcommand{\rho}{P}
\newcommand{\nqs}{({\rm nqs})}

\section{Derivation of Main Result [Eq.~(\ref{e:invariance of KNST})]} \label{app}
\newcommand{\Dt}{\Delta t}

\begin{figure}[b]
 \begin{center}
  \includegraphics[width=.8\hsize]{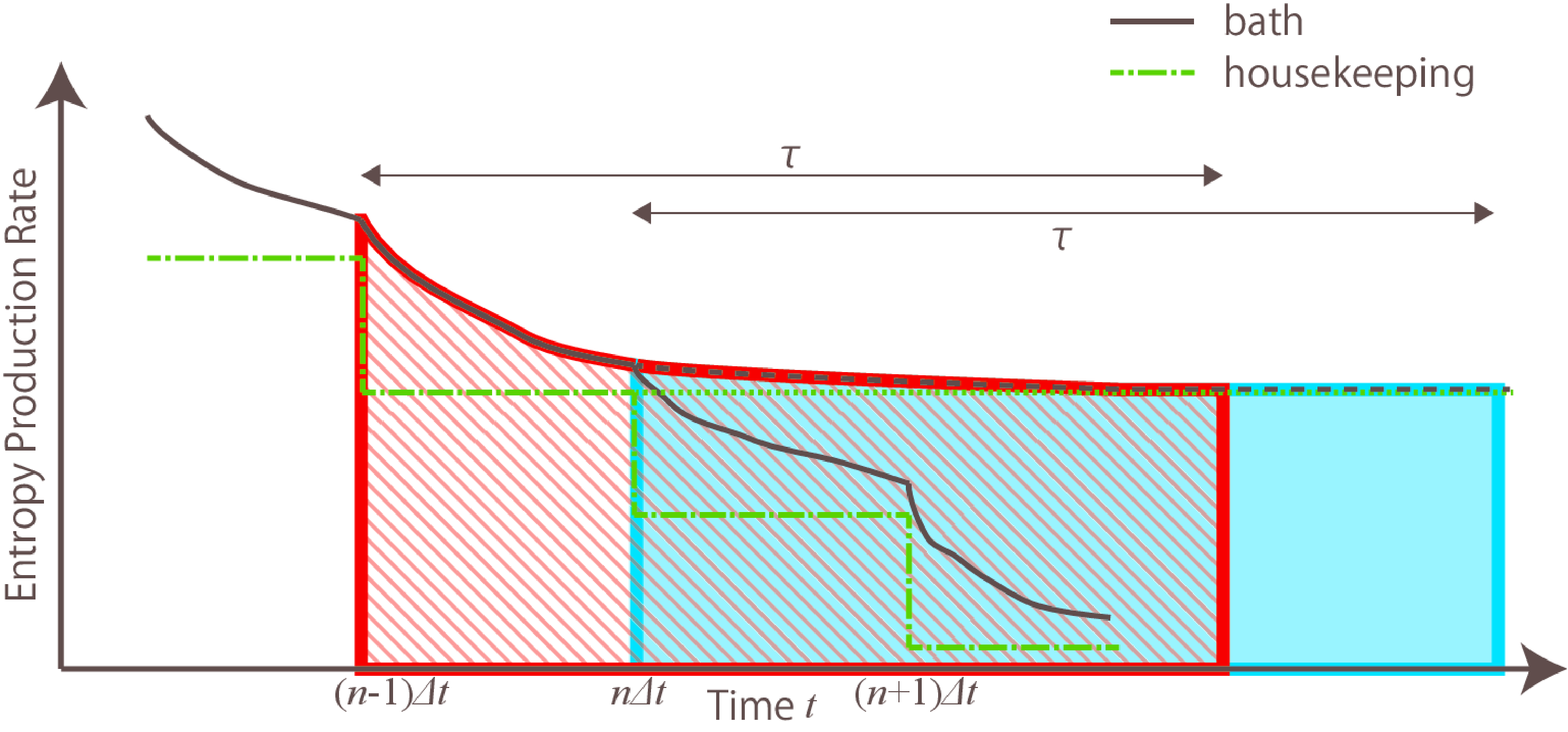}
 \end{center}
 \caption{(Color online) Schematic picture of Eq.~(\ref{e:excess in step-protocol}). Solid (black) and dashed-dotted (green) lines represent the entropy production rate in the baths and its housekeeping part, respectively. Dashed and dotted lines are the corresponding entropy production rates when the controllable parameter is fixed at \(\lt[n\Dt]\).
 Hatched (red) and shaded (blue) areas amount to first and second terms in the last line of Eq.~(\ref{e:excess in step-protocol}). By choosing \(\tau\) sufficiently long compared with the time scale of relaxation to steady states, the difference between two areas gives the excess entropy production on the left hand side of Eq.~(\ref{e:excess in step-protocol}).}
 \label{f:excess in step-protocol}
\end{figure}

First, 
we rewrite the excess entropy production 
in terms of 
the entropy production conditioned by \(\Gamma\) at the initial time point, \(\cTheta[]{\lambda}\) \cite{komatsu_entropy_2010}.
\(\cTheta[]{\lambda}\) is the entropy production in the heat baths during the time interval \([0,\tau]\) when the system is initially set to be \(\Gamma\) and the controllable parameter is fixed at \(\lt[]\).
Here, \(\tau\) should be set sufficiently longer than the relaxation time of the system.
In a \(N\)-step piecewise constant protocol \((\lt^{(N)})_{t\in[0,\mathcal{T}]}\) introduced to approximate the protocol \(\left(\lt\right)_{t\in[0,\mathcal{T}]}\):
\begin{align}
 \lt^{(N)} &= 
 \begin{cases}
 \lt[0] & t = 0
 \\
 \lt[j \Dt] & t \in \left((j-1) \Dt, j \Dt\right] \mbox{ for } j = 1, \ldots, N 
 \end{cases}
 \label{e:piecewise}
\end{align}
the excess entropy production in the heat baths during the time interval \([(n-1)\Dt, n\Dt]\) is given as,
\begin{align}
 &\int_{(n-1)\Dt}^{n\Dt} \sKNSTex[\lt^{(N)}]{\Pm[\Gamma]} \mathrm dt 
 \nonumber \\
 =& \int [\Pt[(n-1)\Dt](\Gamma) - \Pss{\lt[n\Dt]}{\Gamma}] \cTheta[]{\lt[n\Dt]} \mathrm d\Gamma
 \nonumber \\
 & - \int [\Pt[n\Dt](\Gamma) - \Pss{\lt[n\Dt]}{\Gamma}] \cTheta[]{\lt[n\Dt]} \mathrm d\Gamma
 \nonumber \\
 =& \int [\Pt[(n-1)\Dt](\Gamma) - \Pt[n\Dt](\Gamma)] \cTheta[]{\lt[n\Dt]} \mathrm d\Gamma
 .
 \label{e:excess in step-protocol}
\end{align}
We depict the situation of Eq.~(\ref{e:excess in step-protocol}) in \figurename~\ref{f:excess in step-protocol}.
Since \((\lt^{(N)})_{t\in[0,\mathcal{T}]}\) reproduces \(\left(\lt\right)_{t\in[0,\mathcal{T}]}\)
in the limit of \(N\rightarrow\infty\) (\(\Dt := \mathcal{T} / N \rightarrow 0\)),
the excess entropy production in the heat baths during the time interval \([0, \mathcal{T}]\) in the original protocol \(\left(\lt\right)_{t\in[0,\mathcal{T}]}\) may be transformed into,
\begin{align}
 \int_0^\mathcal{T} \sKNSTex[\lt]{\Pm[\Gamma]} \mathrm dt 
 =& -\int_0^\mathcal{T} \mathrm dt \int \rd{\Pt[t](\Gamma)}{t} \cTheta[]{\lt[t]} \mathrm d\Gamma
 .
 \label{e:expression of excess}
\end{align}
Note that since it is natural to expect Eq.~(\ref{e:expression of excess}) to hold irrespective of its derivation, 
the stepwise protocol assumed here does not contradict with the assumption \(\tau_y \ll \tau_\lambda\).

Next, we transform the difference in the symmetrized Shannon entropy increment between the scales of description.
When the time scales are sufficiently separated [i.e., when Eq.~(\ref{e:slaving}) holds], it may be decomposed as,
\begin{align}
 &\{\Hs[\rho_{\mathcal{T}}(x,y)] - \Hs[\rho_{0}(x,y)]\}
 - \{\Hs[\rho_{\mathcal{T}}(x)]
 \nonumber \\
 &- \Hs[\rho_{0}(x)]\}
 \nonumber \\
 =& - \int_0^\mathcal{T} \mathrm dt \frac{\mathrm d}{\mathrm dt} \left[\int \mathrm dx \mathrm dy P_t(x,y) \frac{1}{2} \ln \frac{P_t(x,y)}{P_t(x)} \right.
 \nonumber \\
 & \qquad\qquad + \left. \int \mathrm dx \mathrm dy P_t(x,y) \frac{1}{2} \ln \frac{P_t(x^*,y^*)}{P_t(x^*)} \right]
 \nonumber \\
 =& \int_0^\mathcal{T} \mathrm dt \left[- \int \mathrm dx \mathrm dy \rd{\rho_{{t}}(x,y)}{t} \ln \rho_{\rm ss}^{\lt[t]}(y^*|x^*) \right.
 \nonumber \\
  &+ \frac{1}{2} \int \mathrm dx \mathrm dy \rd{\rho_{{t}}(x,y)}{t} \left\{ \ln \rho_{\rm ss}^{\lt[t]}(y^*|x^*) -  \ln \rho_{\rm ss}^{\lt[t]}(y|x) \right\}
 \nonumber \\
  &- \frac{1}{2} \int \mathrm dx \mathrm dy \rho_{t}(x,y) \dt{\lt}\rd{\ln \rho_{\rm ss}^{\lt[t]}(y|x)}{\lt}
 \nonumber \\
  &\left. - \frac{1}{2} \int \mathrm dx \mathrm dy \rho_{t}(x,y) \dt{\lt}\rd{\ln \rho_{\rm ss}^{\lt[t]}(y^*|x^*)}{\lt}
 + O(\tsratio)\right]
 .
 \label{e:decomposition}
\end{align}
By considering the {\it linear response formula} \cite{komatsu_entropy_2010} which relates the steady state probability density function, \(\Pss{\lt}{\Gamma}\), with \(\cTheta[]{\lambda}\) as,
\begin{align}
 \ln \rho_{\rm ss}^{\lambda}(\X) = -\mathcal{S}(\lambda) - \cTheta{\lambda} + O(\noneq^2)
\end{align}
[\(\mathcal{S}(\lambda)\) is determined from the normalization],
the integrand of first term in the last line of Eq.~(\ref{e:decomposition}) gives the difference in the excess entropy production rate between two scales (and some residual terms):
\begin{align}
 &\int \mathrm dx \mathrm dy \rd{\rho_{{t}}(x,y)}{t} \left[\cSbath{(x,y)}{\lt[t]} - \cSbath{x}{\lt[t]} + O(\noneq^2)\right]
 \nonumber \\
 = 
 &- \sKNSTex[\lt]{\Pm[x,y]}
 + \sKNSTex[\lt]{\Pm[x]}
 + \int \mathrm dx \mathrm dy \rd{\rho_{{t}}(x,y)}{t} O(\noneq^2)
 \nonumber \\
 = 
 &- \sKNSTex[\lt]{\Pm[x,y]}
 + \sKNSTex[\lt]{\Pm[x]}
 \nonumber \\
 &+ \int \mathrm dx \mathrm dy \left\{ \left[\rd{\rho_{{t}}(x)}{t} - \dt{\lt} \rd{\rho_{\rm ss}^{\lt}(x)}{\lt}\right] \rho_{\rm ss}^{\lt}(y|x) \right.
 \nonumber \\
 &\left.+ \dt{\lt} \rd{\rho_{\rm ss}^{\lt}(x)}{\lt} \rho_{\rm ss}^{\lt}(y|x) + P_t(x) \dt{\lt} \rd{\rho_{\rm ss}^{\lt}(y|x)}{\lt}\right\} O(\noneq^2)
 \nonumber \\
 = 
 &- \sKNSTex[\lt]{\Pm[x,y]}
 + \sKNSTex[\lt]{\Pm[x]} 
 + \dt{\lt} O(\noneq^2)
 \nonumber \\
 &+ \int \mathrm dx \mathrm dy \left[\rd{\rho_{{t}}(x)}{t} - \dt{\lt} \rd{\rho_{\rm ss}^{\lt}(x)}{\lt}\right] \rho_{\rm ss}^{\lt}(y|x) O(\noneq^2)
 .
\end{align}
Here, we omitted \(O(\eta)\) term.
Because it may be shown that the time integral of third term gives \(O(\noneq^3)\) contribution for appropriate protocols in the context of extended Clausius relation [e.g. the case where \(\displaystyle \int \mathrm dt \left|\dot{\lt}\right| = O(\noneq)\)], we include it in \(\res[\rm qs]{\Pt[0]{(x)}, \lt}\).
The integrands of second and fourth terms in the last line of Eq.~(\ref{e:decomposition}) are transformed to
\begin{widetext}
\begin{align}
 &\frac{1}{2} \int \mathrm dx \mathrm dy \rd{\rho_{{t}}(x,y)}{t} \left[ \frac{\rho_{\rm ss}^{\lt[t]}(y^*|x^*) - \rho_{\rm ss}^{\lt[t]}(y|x)}{\rho_{\rm ss}^{\lt[t]}(y|x)} + O(\epsilon^2)\right]
 - \frac{1}{2} \int \mathrm dx \mathrm dy \rho_{t}(x,y) \dt{\lt} \frac{1}{\rho_{\rm ss}^{\lt}(y^*|x^*)} \rd{\rho_{\rm ss}^{\lt}(y^*|x^*)}{\lt}
 \nonumber \\
 &= \frac{1}{2} \int \mathrm dx \mathrm dy \left[ \rd{\rho_{{t}}(x)}{t}  \rho_{\rm ss}^{\lt}(y|x) +  P_t(x) \dt{\lt} \rd{\rho_{\rm ss}^{\lt}(y|x)}{\lt}\right]
 \left[ \frac{\rho_{\rm ss}^{\lt[t]}(y^*|x^*)}{\rho_{\rm ss}^{\lt[t]}(y|x)} + O(\epsilon^2)\right]
 \nonumber \\
 &- \frac{1}{2} \int \mathrm dx \mathrm dy \rho_{t}(x^*) \rho_{\rm ss}^{\lt}(y^*|x^*) \dt{\lt} \frac{1}{\rho_{\rm ss}^{\lt}(y|x)} \rd{\rho_{\rm ss}^{\lt}(y|x)}{\lt}
 \nonumber \\
 &= \frac{1}{2} \int \mathrm dx \mathrm dy [\rho_t(x) - \rho_{t}(x^*)] \dt{\lt} \rd{\rho_{\rm ss}^{\lt}(y|x)}{\lt} \frac{ \rho_{\rm ss}^{\lt}(y^*|x^*)}{\rho_{\rm ss}^{\lt}(y|x)}
 + \frac{1}{2} \int \mathrm dx \mathrm dy \left[ \rd{\rho_{{t}}(x)}{t}  \rho_{\rm ss}^{\lt}(y|x) +  P_t(x) \dt{\lt} \rd{\rho_{\rm ss}^{\lt}(y|x)}{\lt}\right] O(\noneq^2)
 \nonumber \\
 &= \frac{1}{2} \int \mathrm dx \mathrm dy [\rho_t(x) - \rho_{t}(x^*)] [{ \rho_{\rm ss}^{\lt}(y^*|x^*) - \rho_{\rm ss}^{\lt}(y|x)}]\dt{\lt} \rd{\ln \rho_{\rm ss}^{\lt}(y|x)}{\lt} 
 \nonumber \\
 &+ \frac{1}{2} \int \mathrm dx \mathrm dy \left[ \rd{\rho_{{t}}(x)}{t} - \dt{\lt} \rd{\rho_{\rm ss}^{\lt}(x)}{\lt}\right] \rho_{\rm ss}^{\lt}(y|x) O(\noneq^2) + \frac{1}{2} \int \mathrm dx \mathrm dy \left[\rd{\rho_{\rm ss}^{\lt}(x)}{\lt} \rho_{\rm ss}^{\lt}(y|x) + \rho_t(x) \rd{\rho_{\rm ss}^{\lt}(y|x)}{\lt}\right]\dt{\lt} O(\noneq^2)
\end{align}
\end{widetext}
Once again, we include the last integral in \(\res[\rm qs]{\Pt[0]{(x)}, \lt}\).
The third term in the last line of Eq.~(\ref{e:decomposition}) becomes \(0\).
Therefore, we obtain the explicit expression of Eq.~(\ref{e:invariance of KNST})
\begin{align}
 \Omega[\Pt[0](x,y), \lt] =& \Omega[\Pt[0](x), \lt]
 \nonumber \\
 &+ \res[\rm qs]{\Pt[0]{(x)}, \lt} + \res[\rm nqs]{\Pt[0]{(x)}, \lt}
 + O(\tsratio)
 \nonumber \\
 &+ \frac{1}{2} \int_0^\mathcal{T} \mathrm dt\int \mathrm dx \mathrm dy \Big\{ [\rho_{t}(x) - \rho_{t}(x^*)]
 \nonumber \\
 &[\rho_{\rm ss}^{\lt[t]}(y^*|x^*) - {\rho_{\rm ss}^{\lt[t]}(y|x)}]
 \frac{\mathrm d\lt}{\mathrm dt} \rd{}{\lambda}\ln \rho_{\rm ss}^{\lambda}(y|x) \Big\}
 ,
 \label{e:explicit invariance of KNST}
\end{align}
where we identify the terms proportional to \(\rd{\rho_{{t}}(x)}{t} - \dt{\lt} \rd{\rho_{\rm ss}^{\lt}(x)}{\lt}\) as an accumulated change of the deviation of the coarse-grained probability density function from the steady state one: \(\res[\rm nqs]{\Pt[0]{(x)}, \lt}\).
The last term of Eq.~(\ref{e:explicit invariance of KNST}) becomes \(O(\noneq^3)\) by imposing Eq.~(\ref{e:symmetry condition}) or Eq.~(\ref{e:overdamped condition}).
\end{document}